\documentclass{article}
\usepackage{spconf,amsmath,graphicx}
\usepackage{glossaries}
\usepackage{balance}
\usepackage[mathscr]{euscript}
\usepackage[dvipsnames]{xcolor}
\usepackage{url}
\usepackage{tabularx,ragged2e}
\usepackage{booktabs}
\usepackage{multirow}
\usepackage[font=footnotesize]{caption}

\def\Ia{\mathbf{I_{\mathsf{a}}}}
\def\Txab{\mathscr{T_{\mathsf{ab}}}}  
\def\Ib{\mathbf{I_{\mathsf{b}}}}
\def\Tab{\mathbf{T_{\mathsf{a b}}}}  
\def\sab{s_{\mathsf{a b}}}
\def\I{{\mathbf I}}

\def\K{{\mathbf K}}
\def\W{{\mathbf W}}

\def\Kiv{{{\K}}}  

\def\F{\mathbf{F}}
\def\c{\mathbf{c}}

\def\LP{\dutchcal{LP}}

\def\CR{\mathcal{C}}
\def\TR{\mathcal{A}}
\def\SR{\mathcal{S}}

\def\Txf{\mathscr{T}_{\hat{s} \hat{\alpha} \hat{\c}}}
\def\Pfm{P_{\MFMT}}
\def\phi{{\boldsymbol \Phi}}
\def\alphaab{\alpha_{\mathsf{a b}}} 
\def\cabx{c_{\mathsf{ab}_x}} 
\def\caby{c_{\mathsf{ab}_y}} 

\def\x{\mathbf{x}}
\def\LP{\mathrm{LP}}
\def\FMT{\mathrm{FM}}
\def\MFMT{\mathrm{MFM}}

\newacronym{prnu}{PRNU}{Photo Response Non-Uniformity}
\newacronym{spn}{SPN}{Sensor Pattern Noise}
\newacronym{ncc}{NCC}{Normalized Cross-Correlation}
\newacronym{dft}{DFT}{Discrete  Fourier Transform}
\newacronym{pce}{PCE}{Peak-to-Correlation Energy}
\newacronym{fm}{FM}{Fourier-Mellin transform}
\newacronym{mfm}{MFM}{modified Fourier-Mellin transform}

\title{A Modified Fourier-Mellin Approach for \\Source Device Identification on Stabilized Videos}
\name{Sara Mandelli$^*$,  Fabrizio Argenti$^\dag$, Paolo Bestagini$^*$, Massimo Iuliani$^\dag$$^\ddag$, Alessandro Piva$^\dag$$^\ddag$, Stefano Tubaro$^*$
	\thanks{This material is based on research sponsored by DARPA and AFRL under agreement number FA8750-16-2-0173. The U.S. Government is authorized to reproduce and distribute reprints for Governmental purposes notwithstanding any copyright notation thereon. The views and conclusions contained herein are those of the authors and should not be interpreted as necessarily representing the official policies or endorsements, either expressed or implied, of DARPA and AFRL or the U.S. Government. This work was supported by the PREMIER project, funded by the Italian Ministry of Education, University, and Research within the PRIN 2017 program.}}
\address{$^*$Dipartimento di Elettronica, Informazione e Bioingegneria, Politecnico di Milano - Milan, Italy \\
	$^\dag$Department of Information Engineering, University of Florence - Florence, Italy\\
	$^\ddag$FORLAB - Multimedia Forensics Laboratory - Prato, Italy}
\begin{document}
\ninept
\maketitle
\begin{abstract}
To decide whether a digital video has been captured by a given device, multimedia forensic tools usually exploit characteristic noise traces left by the camera sensor on the acquired frames.
This analysis requires that the noise pattern characterizing the camera and the noise pattern extracted from video frames under analysis are geometrically aligned.
However, in many practical scenarios this does not occur, thus a re-alignment or synchronization has to be performed.
Current solutions often require time consuming search of the re-alignment transformation parameters.
In this paper, we propose to overcome this limitation by searching scaling and rotation parameters in the frequency domain.
The proposed algorithm tested on real videos from a well-known state-of-the-art dataset shows promising results.

\end{abstract}
%

\section{Introduction}
\label{sec:intro}
Multimedia forensics keeps developing technologies to identify the camera originating a digital image or a digital video. 
Currently, the most promising technique
 is based on the analysis of \gls{spn} or \gls{prnu}, left by the acquisition device into the visual content.
This trace is useful to identify the video source since it is universal (i.e., every camera sensor introduces one) and unique (i.e., \gls{prnu} from two different sensors  are uncorrelated)~\cite{lukas2006digital, mondaini2007detection}.
Moreover, \gls{prnu}  has proved to be significantly robust to commonly used processing, like JPEG compression~\cite{lukas2006digital}, or uploading to social media platforms~\cite{castiglione2013experimentations,  bertini2016social}.

\gls{prnu}-based source identification process consists in verifying the match between a query image or video frame and a fingerprint characterizing a reference camera.
The strategy involves two main steps:
i) a reference fingerprint is derived from still images or videos acquired by the source device;
ii) the query fingerprint is estimated from the investigated content and then compared with the reference to verify the possible match, in form of a correlation.
If the query content was acquired by the reference camera, then a high correlation is expected.

The previous scheme works under the hypothesis of perfect geometrical alignment between the reference and test fingerprints.
If a geometrical transformation is applied to the query content, a pixel grid misalignment between the query and the reference fingerprint arises, thus hindering the detection.
Such a case occurs in multiple scenarios:
when an image or a video has been acquired with different resolution settings or it is cropped and resized due to the upload in a social media;
if a malicious user slightly distorts the content to remove the sensor traces;
when a query video is tested against a reference estimated from still images;
when a video has been created in presence of electronic image stabilization.
In all these cases, the \gls{prnu}  extracted from the query is misaligned with the reference fingerprint, and thus a geometric re-synchronization between them has to be carried out before the matching operation.

The first solution to this problem was proposed in~\cite{goljan2008camera}, where the case of cropped and downscaled images was studied.
The authors show that it is possible to parameterize the \gls{ncc}  between the reference fingerprint and the query noise with respect to the scaling factor.
The \gls{ncc} peak position for a given scaling factor provides an estimate of the shift.
While the \gls{ncc} can be efficiently computed in the Fourier domain, a brute force search is needed to determine the scaling factor.
By following the same rationale, more recent papers \cite{taspinar2016source, iuliani2019hybrid, Mandelli2019} extend the proposed methodology also considering rotation and the more challenging scenario of video analysis.
As a matter of fact, modern acquisition pipelines usually include electronic stabilization that undermines \gls{prnu}-based attribution technique. In these cases, \gls{prnu}-based techniques only work if geometric transformations are properly estimated and compensated for, which is a computational complex operation.

In this paper, we focus on the problem of camera attribution of stabilized video sequences based on \gls{prnu}.
Specifically, we propose a method to align frame fingerprints with the reference \gls{prnu} 
by recovering the scaling, shift and rotation parameters introduced by electronic stabilization. 
We overcome the problem of computational complexity by searching for scaling and rotation parameters in the frequency domain thanks to a modified version of the \gls{fm}.
Results obtained on the well known Vision dataset \cite{shullani2017vision} show that the proposed method provides extremely efficient results whenever rotation and scaling operations are applied to video frames.
When also shift is taken into account, the gain compared against the state-of-the-art \cite{Mandelli2019} depends on the video content.


\section{Background and problem statement}
\label{sec:problem}
In this section we introduce the background on Fourier-Mellin (FM) transform and define the problem we are tackling in this paper.

\vspace{.5em}\textbf{Fourier-Mellin Transform.}
\label{subsec:fm_theory}
The \gls{fm} transform enables to estimate scale, rotation and shift transformations between two images in closed form \cite{reddy1996fft}.

Given an image $\I$, the \gls{fm} transform is expressed as the log-polar mapping of the magnitude of the image Fourier transform, i.e.,
\begin{equation}
\FMT\{\I\} = \LP \{|\F| \},
\end{equation}
where $ \LP \{ \cdot \} $ is the operator computing the log-polar mapping, and $|\F|$ is the magnitude of the Fourier transform.

Let us consider two images $\Ia$ and $\Ib$ that are linked through a similarity transformation, i.e.,  $\Ia = \Txab \lbrace\Ib\rbrace$, where $\Txab$ applies the transformation identified by the matrix
\begin{equation}
\Tab= \left[\begin{matrix}
\sab \cdot \cos \alphaab, & -\sab \cdot \sin \alphaab, &\cabx\\ 
\sab \cdot \sin \alphaab, & \sab \cdot \cos \alphaab, &\caby
\end{matrix}\right],
\end{equation}
where $\sab$ represents scaling, $\alphaab$ rotation and $\c = [\cabx, \caby]$ horizontal and vertical shift.
In this scenario, it is possible to show that $\FMT\{ \Ia \}$ is a shifted version of $\FMT\{ \Ib \}$.
More formally,
\begin{equation}
\FMT\{ \Ia \}(\rho, \alpha) = \FMT\{ \Ib \}(\rho - \log \sab, \alpha - \alphaab), 
\label{eq:02_logpolar_theory_magnitude}
\end{equation}
where $ \rho $ is the radial coordinate and $ \alpha $ the rotational coordinate.
It is therefore possible to estimate scale $\sab$ and rotation $\alphaab$ by looking at the peak position of the phase correlation function between $\FMT\{ \Ia \}$ and $\FMT\{ \Ib \}$ independently from shift \cite{reddy1996fft}.
Once $\sab$ and $\alphaab$ are estimated, the two images can be realigned apart from translation.
The relative shift can then be estimated by looking at the peak position of the phase correlation computed between the two realigned images in the pixel domain \cite{reddy1996fft}.

\vspace{.5em}\textbf{Problem formulation.}
\gls{prnu} is typically modeled as a multiplicative noise pattern introduced by any device in all acquired images or videos \cite{lukas2006digital, chen2008determining}.
In the field of forensics analysis, it is well known that \gls{prnu} can be exploited for inferring whether an image was shot by a certain device.
For instance, given a test image $\I$ and a device PRNU $ \K$, we can compute the \gls{pce} between the noise residual $\W$ extracted from the image and the \gls{prnu} pixel-wise scaled by $\I$, i.e., $\textrm{PCE}(\W, \K \cdot \I )$. 
Indeed, $\textrm{PCE}$ measures the correlation between the noise traces left on $\I$ and the device \gls{prnu} independently of potential shift misalignment, as the correlation peak is searched over all possible mutual shifts between them.
If the \gls{pce} is greater than a confidence threshold, we attribute $\I$ to the device \cite{lukas2006digital, chen2008determining}.

The extension of \gls{prnu}-based strategies for attributing video frames to a specific device suffers from some issues due, for instance, to higher compression rates and lower pixel resolutions.
As a matter of fact, the previously described \gls{pce} test cannot be directly performed, being the \gls{prnu} resolution typically higher than the size of the recorded video frames.
Moreover, in-camera video stabilization techniques, which are now becoming one of the must-have device specifications, strongly hinder the traces left by \gls{prnu}, as video frames may be warped by means of geometrical transformations (e.g., cropping, rotation, scaling, etc.) in order to generate a stable video sequence \cite{Mandelli2019, Grundmann2018}.
As a consequence, the attribution of a video frame to a specific device can represent a much more challenging task than common image-camera attribution.

In this paper, we exploit \gls{prnu}-based traces to investigate the problem of device attribution when testing in-camera stabilized video frames.
Specifically, given a device fingerprint $ \Kiv $ and a frame $\I$ coming from a stabilized video sequence, we aim at exploiting the \gls{prnu} traces left on $\I$ in order to detect whether it has been recorded by the analyzed device.
To do so, we assume that geometric transformations can be approximated by similarities \cite{iuliani2019hybrid, Mandelli2019} and we propose a geometrical realignment strategy based on a modified version of \gls{fm} transform applied to both the device fingerprint and the frame noise residual.
Specifically, the proposed \gls{mfm} enables comparing a device fingerprint and a noise residual independently from scaling and rotation operations.
The next section provides all the details of the proposed method.
\section{Proposed method}
\label{sec:method}
In order to attribute a video frame $\I$ to a device whose reference fingerprint is $\Kiv$, we follow a pipeline based on a few steps:
(i) noise extraction;
(ii) geometric transformation estimation;
(iii) geometric compensation and matching.
In the following, we illustrate all the steps of the pipeline.

\vspace{.5em}\textbf{Noise extraction.} 
As in the common \gls{prnu}-based attribution algorithm, we extract the noise residual $\W$ from frame $\I$.
This is done using the strategy proposed in \cite{lukas2006digital, chen2008determining}:
(i) the noise is extracted through wavelet-based denoising;
(ii) a series of post-processing steps (e.g., zero-averaging rows and columns, Wiener filtering, etc.) are applied to further enhance the noise residual $\W$.

\vspace{.5em}\textbf{Geometric transformation estimation.}
In order to match $\W$ and the fingerprint $\Kiv$, we first need to search for the geometrical transformation that might link them.
In principle, assuming that video frames warping can be approximated by a similarity transformation \cite{iuliani2019hybrid, Mandelli2019}, aligning a noise residual and a reference device fingerprint by means of Fourier-Mellin may seem straightforward.
In practice, differently from the Fourier-Mellin theory presented in Section~\ref{subsec:fm_theory}, the two terms to compare (i.e., $\W$ and $\Kiv$) are not exactly one the transformed version (by means of a similarity transformation) of the other.
First, the geometric transformation introduced by stabilization is not necessarily a similarity, but can include perspective distortions (on the entire frame or a localized portion of it) as well \cite{Grundmann2018, Wang2018}. 
Second, the noise residuals of video frames may contain scene content and noise contributions which are not present in the reference device fingerprint.

The primary consequence of this dissimilarity is that selecting only the Fourier magnitudes for estimating scale and rotation between the two terms, as reported in \eqref{eq:02_logpolar_theory_magnitude}, may be not precise. 
Indeed, we verified that phase correlation between $\FMT \{ \Kiv \}$ and $\FMT \{\W\}$ does not show a pronounced peak, thus leading to a strongly hindered estimation of scale, rotation and shift.
In order to overcome this issue we modify the Fourier-Mellin pipeline in two ways.

First, we propose to embed the phase term of the Fourier transform in addition to the magnitude to the Fourier-Mellin pipeline.
The modified Fourier-Mellin transform of $\I$ can be thus defined as:
\begin{equation}
\MFMT\{\I\} = \LP \{\F\}, 
\end{equation}
where $\LP \{\F\} $ is the log-polar mapping of the image Fourier transform (including magnitude and phase).
On one hand, phase adds more information, which is very useful for angle and scale estimation.
On the other hand, this operation comes with a cost.
The natural drawback of this approach is that we cannot isolate anymore the estimation of scale and rotation from the estimation of the shift. 
Indeed, in this case, phase correlation does not exclusively depend on scale and rotation transformations, but also on translation between the two terms. 
The Fourier-Mellin pipeline works only if $ \W $ and $\Kiv$ are almost perfectly aligned in terms of translation, i.e., if their mutual shift is basically $ 0 $ pixels in both horizontal and vertical directions. 
In other words, including the Fourier phase term, we first have to correctly realign the \gls{prnu} traces left on the noise residual with those on the reference fingerprint for what concerns the relative shift, then we can convert the Fourier transforms into log-polar domain and estimate the remaining parameters.

The second proposed modification helps enabling faster computations.
It has been shown that a properly selected portion of the \gls{prnu} frequency spectrum can be sufficient to achieve good attribution performance (e.g., through subsampling \cite{bondi2019improving}).
In this vein, notice that a 2D frequency band becomes a rectangular band if the frequency spectrum is converted in log-polar domain.
We propose to literally cut the frequency content of $\Kiv$ and $\W$ by cropping the log-polar Fourier transform of $\Delta_{\rho}$ samples along the $\rho$ dimension.
The cropping center corresponds to the coordinate of the highest energy peak of $\MFMT \{ \K \}$ evaluated as a function of $\rho$.
Despite this step might seem irrelevant, this \textit{strongly} reduces the amount of frequency samples to be correlated, thus lowering the computational cost.
We define the modified Fourier-Mellin transform followed by cropping as:
\begin{equation}
\MFMT_{\Delta_{\rho}}\{\I\} = [\LP \{\F\} ]_{\Delta_{\rho}}.
\end{equation}

\begin{figure}[t]
	\centering
	\includegraphics[width=\columnwidth]{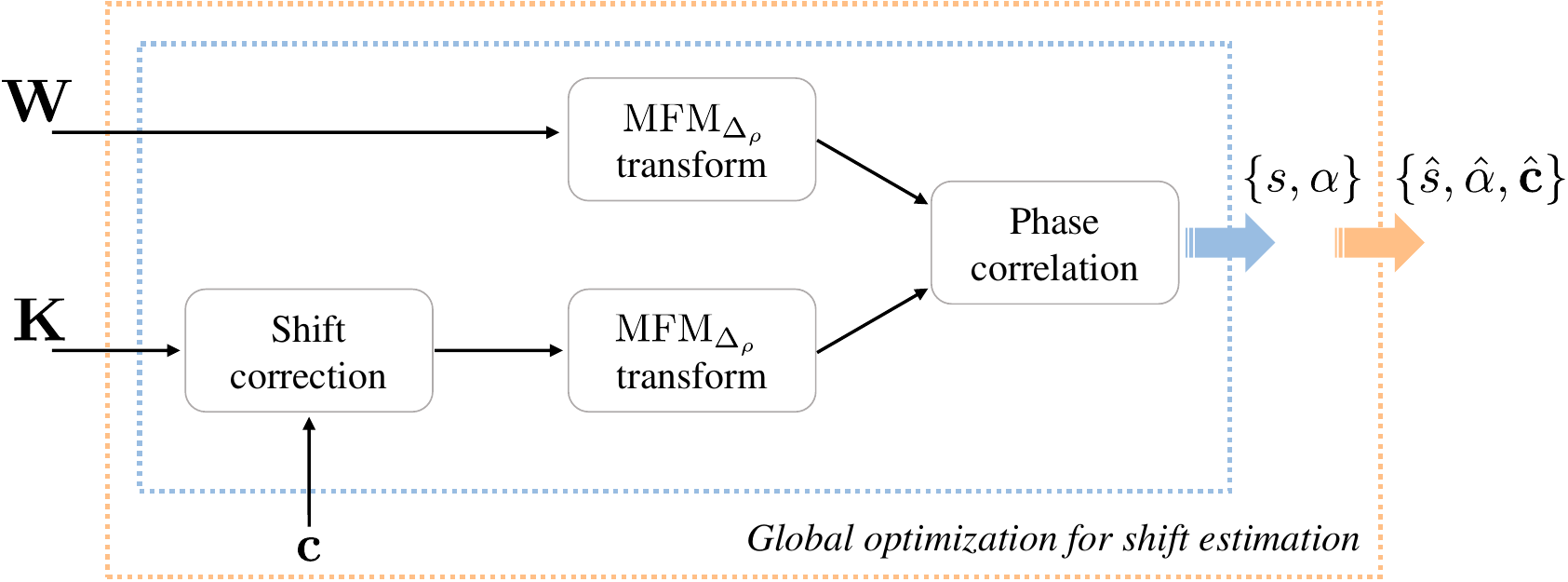}
	\caption{Scheme of the proposed method for similarity estimation between noise residual $\W$ and reference device fingerprint $\Kiv$. The global optimizer searches for shift candidates, while the proposed $\MFMT_{\Delta_{\rho}}$ transform provides an estimate of scale and rotation for each shift.}
	\vspace{-1.3em}
	\label{fig:fm_pipeline}
\end{figure}

By considering the added phase term and the frequency cropping step, the best similarity parameters can be estimated solving a maximization problem. 
Formally, 
\begin{equation}
\hat{s}, \hat{\alpha}, \hat{\c} =\underset{s \in \SR,\alpha \in \TR, \c \in \CR}{\arg \max}  \phi [ \MFMT_{\Delta_{\rho}} \{\W\},  \MFMT_{\Delta_{\rho}} \{\Kiv(\x - \c) \} ], 
\label{eq:fm_max_problem}
\end{equation}
where $\phi$ represents the phase correlation and vector $\x$ refers to horizontal and vertical pixel coordinates.

Notice that, for each shift candidate value, scale and rotation parameters can be very quickly estimated in closed form through phase correlation. 
Therefore, we only need to optimize over different shift values.
However, 
gradient descent strategies to solve \eqref{eq:fm_max_problem} suffer from the non-convex behavior of phase correlation as a function of the shift.
Especially in video sequences characterized by outdoor scenarios or user motion, the actual peak value can be hard to find with gradient descent algorithms.
The maximization problem as a function of the shift can be solved by resorting to global optimization techniques.
It is worth noting that the translation between $ \W $ and $\Kiv$ can be assumed with slight approximation to imply integer shift in horizontal and vertical directions, i.e., to represent a certain number of pixels.
We propose to exploit a global optimization algorithm known as genetic algorithm that allows an efficient estimation of integer parameters \cite{matlab_opt}.
In a nutshell, our
method is shown in Fig.~\ref{fig:fm_pipeline}. 

\vspace{.5em}\textbf{Geometric compensation and matching.}
After estimating the similarity transformation $\Txf$, last steps consist in:
(i) applying $\Txf$ to $\Kiv$ in order to realign the \gls{prnu} traces left on $\Kiv$ with those of $\W$;
(ii) resorting to \gls{pce} as strategy for a correct source device identification.
We compute $ \Pfm $ as
\begin{equation}
	\Pfm = \textrm{PCE} (\W,  \Txf(\Kiv)).
	\label{eq:fm_pce_test}
\end{equation}
As in standard \gls{prnu} attribution tests, by thresholding  $ \Pfm $ it is possible to detect whether the frame under analysis belongs to the tested device.
In case multiple frames are available, it is possible to repeat the whole procedure and fuse the results obtained with different frames (e.g., maximum \gls{pce} picking, majority voting, etc.).

\section{Experimental analysis}
\label{sec:results}
In this section we report all the details about the performed experimental campaign and the achieved results.

\vspace{.5em}\textbf{Dataset.}
Our datasets have been extracted from Vision dataset, which includes both images and videos from $11$ major brand devices \cite{shullani2017vision}.
For building the \gls{prnu} related to each device, we select all the available images taken by the device depicting flat scenes \cite{chen2008determining}. 
Then, each fingerprint $\Kiv$ is built by scaling and cropping the \gls{prnu}, using the image to video warping parameters reported in \cite{Mandelli2019}.
Regarding video frames, we select only devices with Full-HD video resolution (i.e., $ 1920 \times 1080 $ pixels).
For the sake of clarity, we make use of the same device nomenclature presented in \cite{shullani2017vision}, creating two test datasets: a \emph{non-stabilized} dataset, selecting non-stabilized devices D03, D11, D17, D21, D24
from $5$ different brands, and a \emph{stabilized} dataset that includes all the $14$ available stabilized devices.

Notice that the considered video frames contain both static and motion scenes, depicted as \emph{still, panrot, move} in \cite{shullani2017vision}, and can include almost flat content as well as significant texture presence, denoted as \emph{flat, indoor, outdoor} in \cite{shullani2017vision}.
In particular, we only make use of the I-frames, as the \gls{prnu} traces left on them are likely to be more reliable than those left on inter-predicted frames \cite{taspinar2016source, chuang2011exploringa}.
Furthermore, in light of past investigations about the first I-frame of stabilized video sequences, we always discard it from the experiments \cite{Mandelli2019, Grundmann2018}. 

\vspace{.5em}\textbf{\gls{mfm} parameters.}
To compute the $\MFMT$ transform, we evaluate the 2D Fourier transform over $4096 \times 4096$ samples after zero-padding residue and reference fingerprint in the pixel domain, in order not to introduce undesired border effects. Then, we convert both terms into log-polar domain, following the default parameters provided by \cite{matlab_images_logpolar}, ending up with $\MFMT$ transforms having $2896$ $\rho$-samples and $2281$ $\alpha$-samples. 
We verified that the sampling grid for $\rho$ and $\alpha$ dimensions allows a correct estimation of scaling factor and rotation angle.
Eventually, we crop $\MFMT$ transforms along $\rho$ dimension according to the chosen number of samples $\Delta_{\rho}$.

The exploited genetic algorithm mimics biological evolution to find a reliable shift estimation. Precisely, it has the following parameter configuration: a population size of $50$ individuals, which iteratively update the cost function for a maximum of $50$ iterations. Remaining parameters are those defined in \cite{matlab_opt}.

\vspace{.5em}\textbf{Performance in a controlled scenario.}
In order to assess the accuracy in attributing video frames to the correct device, we investigate the proposed method in a controlled scenario.
Specifically, considering the \emph{non-stabilized} dataset, we randomly select $27$ I-frames per device, taking care of equally distributing motion and static scenes, as well as flat and textured content.
We end up with a total amount of $135$ video frames.
In particular, we select only frames which report acceptable \gls{pce} values with the device fingerprint (i.e., \gls{pce} $\geq 60$, as suggested in \cite{taspinar2016source, Mandelli2019}).
Then, we warp each frame by means of a similarity transformation, randomly selecting the parameters from some realistic ranges \cite{Grundmann2018}, namely $\SR = [0.9, 1.1]$, $\TR = [-3, 3] \, \textrm{deg}$, $\CR = [-90, 90] \, \textrm{pixels}$, related to scale, rotation angle, horizontal and vertical shifts, respectively.
We verified these ranges include the vast majority of possible similarity transformations between stabilized video frames and reference fingerprint.

\begin{figure}[t]
	\centering
	\includegraphics[width=\columnwidth]{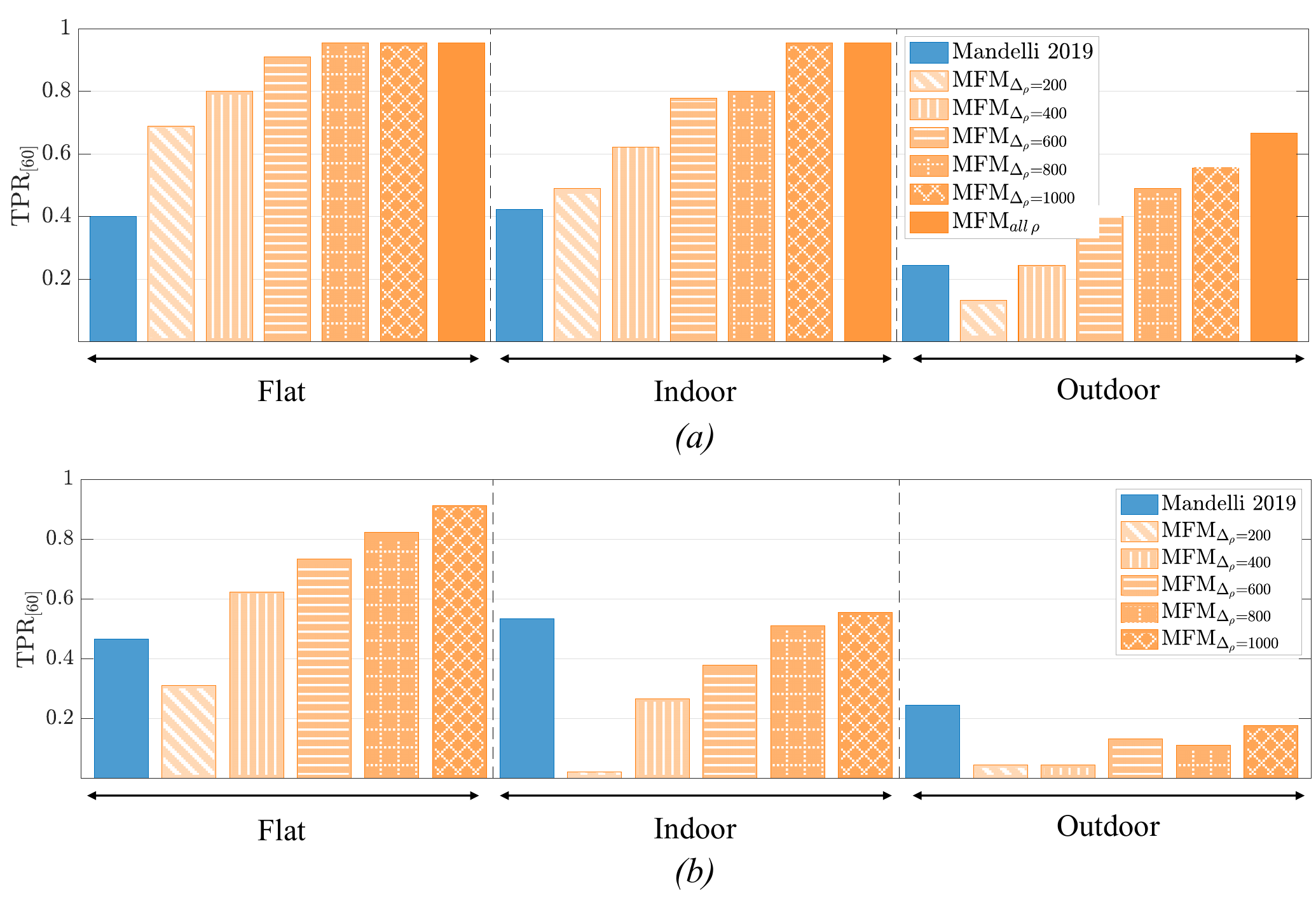}
	\caption{Accuracy on synthetically warped non-stabilized video frames: \textit{(a)} only scale and rotation are applied; \textit{(b)} a complete similarity is applied. The proposed method (orange) can be tuned to used different amount of frequency samples ($\Delta_{\rho}$), thus becoming slower but more accurate.}
	\vspace{-1.3em}
	\label{fig:no_stab}
\end{figure}

We aim at estimating the applied transformation using the proposed strategy, comparing the performance with the method presented in \cite{Mandelli2019}.
Specifically, we exploit the same parameter configuration for the particle swarm strategy \cite{matlab_opt, Kennedy2011} as reported in \cite{Mandelli2019}, which enables to estimate the similarity transformation returning the highest \gls{pce} between $\W$ and $\Kiv$.
For what regards the search bounds of scale and rotation parameters, we suppose these to be known at investigation side, thus they coincide with $\SR$ and $\TR$.
Notice that method \cite{Mandelli2019} does not need to fix bounds for the shift parameters as these can be estimated without the need of optimization.
Following similar considerations, the proposed $\MFMT$ strategy fixes the search range for shift parameters exactly to $\CR$, while scale and rotation do not require optimization.

Computational time and true positive rate evaluated for a PCE threshold of $60$ (i.e., $\mathrm{TPR}_{[60]}$) are the chosen accuracy metrics to compare the two strategies.
The average time for estimating the similarity transformation on each frame with the method \cite{Mandelli2019} is $41 \, \textrm{s}$, while $\MFMT$ strategy changes its temporal requirement depending on $\Delta_{\rho}$ (e.g, using $\Delta_{\rho} = 200$ requires only $21 \, \textrm{s}$ on average). Generally, the required time linearly grows with $\Delta_{\rho}$.

Fig.~\ref{fig:no_stab} shows results as a function of the scene content of video frames (i.e., flat, indoor and outdoor).
Specifically, Fig.~\ref{fig:no_stab}\textit{(a)} reports results where only scale-rotation transformations were applied.
The shift between noise residuals and $\Kiv$ is assumed to be known.
Fig.~\ref{fig:no_stab}\textit{(b)} reports results where a complete similarity transformation has been applied.
It is worth noting that, in case the shift parameter is known and only scale and rotation parameters should be estimated, our proposal can be a viable solution for very fast identification.
Since scale and rotation can be estimated without the need of optimization, the computational time reduces to less than one second. 
The more the selected $\Delta_{\rho}$ samples, the better the accuracy of $\MFMT$ strategy, which overcomes results of \cite{Mandelli2019}. 
Furthermore, in this case there is no need for global optimizers, thus the potential optimization error reduces to zero.
In case \textit{(b)}, $\MFMT$ shows better or basically equivalent results to \cite{Mandelli2019} for flat and indoor scenarios, while outdoor frames seem to be more challenging for the proposed method. 

\vspace{.5em}\textbf{Performance on stabilized videos.}
In order to show the potentiality of $\MFMT$ approach in dealing with source device identification problem on real videos, we apply the proposed method to the stabilized video sequences.
Following previous considerations, we set as search range for the mutual shift $\CR = [-90, 90]$ both in horizontal and vertical directions. 
For clarity's sake, we use the very same accuracy metrics presented in \cite{Mandelli2019}, i.e., the area-under-the-curve $ \mathrm{AUC} $ and $ \mathrm{TPR}_{@0.01} $ of ROC curves, averaged over all devices.
Precisely, $ \mathrm{TPR}_{@0.01} $ corresponds to the rate of correct attributions evaluated when the false positive attribution rate is equal to $0.01$. 

\begin{figure}[t]
	\centering	\includegraphics[width=.7\columnwidth]{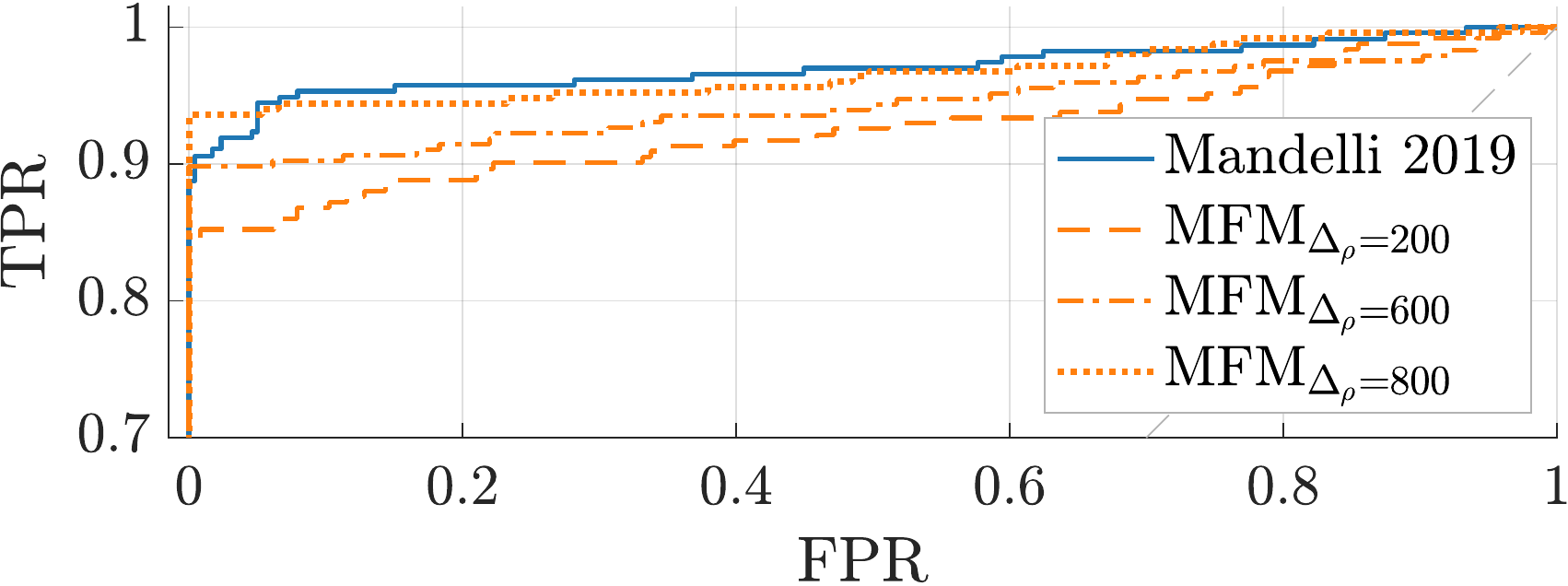}
	\caption{ROC curves obtained testing $ 10 $ I-frames with the proposed strategy, as a function of the number of used frequency samples, i.e., $\Delta_{\rho}$. Results are compared to those of \cite{Mandelli2019} evaluated using $10$ I-frames.}
	\label{fig:fm_roc}
\end{figure}
\begin{table}[t]
	\caption{$ \mathrm{AUC} $ and $ \mathrm{TPR}_{@0.01} $ testing $ 10 $ random I-frames per video query, together with average computational time per frame, evaluated with $\MFMT$ and \cite{Mandelli2019} methods. }
	\label{tab:fm_table}
	\centering
	\def\arraystretch{1.3}
	\resizebox{\columnwidth}{!}{
		\begin{tabular}{cccccc} \toprule[1.5pt]
			$\mathrm{Method}$	  &$\MFMT_{\Delta_{\rho} = 200} $   &$\MFMT_{\Delta_{\rho} = 400} $   &$\MFMT_{\Delta_{\rho} = 600} $  &$\MFMT_{\Delta_{\rho} = 800} $   &\cite{Mandelli2019}  	\\  \midrule[1.5pt]
			$ \mathrm{AUC} $      				& $ 0.93 $				   				& $0.93$										& $ 0.94$                                              & $ 0.97$ 								& $0.97$\\ \midrule[0.1pt]
			$ \mathrm{TPR}_{@0.01} $      & 	$ 0.85 $							& $0.86$									  & $  0.90$                                          & $ 0.94 $ 								& $0.91$ \\ \midrule[0.1pt]
			Time [s]      								&$ 22 $										& $46$									  		& $  69$                                          	& $105$ 								& $57$ \\ \midrule[0.1pt]
		\end{tabular}
	}
\end{table}  
We show the attribution results achieved by testing $ 10 $ random I-frames per video query and picking the maximum value among the computed \gls{pce}s.  
Specifically, we test different values for the number of used frequency samples (i.e., $\Delta_{\rho}$) and always report results achieved by \cite{Mandelli2019} over the same dataset.
Fig.~\ref{fig:fm_roc} draws the ROC curves and Table~\ref{tab:fm_table} depicts the achieved $ \mathrm{AUC} $ and $ \mathrm{TPR}_{@0.01} $ as a function of $\Delta_{\rho}$.
Moreover, last row of Table~\ref{tab:fm_table} reports the average required computational time [seconds] for testing one query frame according to the chosen strategy, considering matching cases as well as non-matching ones.
It is worth noticing that the proposed approach can overcome results of \cite{Mandelli2019}, provided that a sufficient amount of frequency samples is selected. 
Furthermore, the $\MFMT$ strategy enables fast computations as well, at the expense of a slightly reduced accuracy, but still acceptable.

\section{Conclusions}
\label{sec:conclusions}

In this paper, we propose an alternative solution for solving the source device identification problem on stabilized videos. 
Specifically, we re-synchronize video frames and device reference fingerprint by estimating the re-alignment transformation with a modified version of the Fourier-Mellin transform.
In doing so, we search the scaling and rotation parameters in the frequency domain, whereas unknown translations can be estimated leveraging global optimization strategies.
Moreover, we propose to use a reduced amount of Fourier-Mellin transform samples to estimate the warping configuration, thus enabling fast computations.

The experimental campaign is conducted on a publicly available dataset.
Results are promising and show enhanced performance with respect to state-of-the-art. 
This is especially true in situations where only scale and rotation parameters should be estimated: experiments performed in a synthetic set-up reveal that the proposed method can be much faster and accurate than existing methodologies.

\balance
\bibliographystyle{IEEEbib}
\bibliography{biblio}

\end{document}